# SENSITIVITY OF INFERENCES IN FORENSIC GENETICS TO ASSUMPTIONS ABOUT FOUNDING GENES

By Peter J. Green and Julia Mortera[1]

*University of Bristol and Università Roma Tre*

Many forensic genetics problems can be handled using structured systems of discrete variables, for which Bayesian networks offer an appealing practical modeling framework, and allow inferences to be computed by probability propagation methods. However, when standard assumptions are violated—for example, when allele frequencies are unknown, there is identity by descent or the population is heterogeneous—dependence is generated among founding genes, that makes exact calculation of conditional probabilities by propagation methods less straightforward. Here we illustrate different methodologies for assessing sensitivity to assumptions about founders in forensic genetics problems. These include constrained steepest descent, linear fractional programming and representing dependence by structure. We illustrate these methods on several forensic genetics examples involving criminal identification, simple and complex disputed paternity and DNA mixtures.

**1. Introduction.** Forensic genetics is concerned with a variety of inferential problems based on DNA profile data, for example, involving criminal identification or disputed paternity. Inference in these settings is carried out under an assumed probability model, a structured stochastic system of (largely) discrete variables, including typically genes and genotypes of the individuals involved. In such problems, some of the probability distributions specifying the model are not truly known with certainty.

In discrete models in the forensic genetics setting, there are two main aspects of such model uncertainty: (a) assumptions about founders—the default assumption being that all individuals of unknown genotype whose parents are not part of the model are assumed drawn from a homogeneous population in the Hardy–Weinberg equilibrium, with known allele frequencies, and (b) assumptions about mutation—the default being that there is none. Everything else is determined by Mendelian inheritance (granted that the choice of genetic markers used in these problems rules out linkage disequilibrium).

[1]Supported by an Italian PRIN grant.

In this paper we examine the first of these two issues, and study the effect of varying assumptions on the "founding genes," that is, the variables in the model that represent genes inherited from individuals not explicitly represented. Several concrete phenomena:

- unknown allele frequencies,
- identity by descent among founders,
- heterogeneity (the existence of subpopulations),
- lack of the Hardy–Weinberg equilibrium,

that generate correlations between founding genes can be studied by considering the effect of perturbing the joint distribution of the founding genes on the conditional probabilities (posterior inferences) of interest.

Bayesian networks, with inferences computed by probability propagation methods ("junction tree algorithms"), offer an appealing practical modeling framework for structured systems involving discrete variables in numerous domains, including forensic genetics, and we will make extensive use of such networks for model specification and computation of inferences. Other authors have approached related questions of sensitivity using an algebraic approach; see Laurie and Weir (2003), Weir (2007b), Song and Slatkin (2007) among others. However, building on that of Dawid et al. (2002) and subsequent authors, our experience suggests that a Bayesian network approach, although not delivering an explicit algebraic solution, is more flexible and powerful, especially in complex settings.

The outline of the paper is as follows. Brief introductions to the genetic terminology used is given in Section 2 and to Bayesian networks in Section 3.1; the different methodologies for assessing sensitivity are discussed in Section 3. Various forensic identification examples are illustrated in Section 4, and in Section 5, a number of scenarios are discussed that are variations on the baseline assumptions, such as uncertain allele frequencies, identity by descent, population heterogeneity and combinations thereof. Results for the baseline and these scenarios are shown in Section 6, and finally in Section 7 we draw conclusions and discuss further developments.

## 2. Genetic background.

To set the scene, we need some basic facts about DNA profiles; for a more detailed explanation see Butler (2005).

Here we adopt a slightly nonstandard usage of the term *gene*, which for our purpose is simply defined as an identified stretch of DNA and is the entity transmitted from parent to offspring. More precisely, a *gene* is a particular sequence of the four *bases*, represented by the letters A, C, G and T. A specific position on a chromosome is called a *locus* (hence, there are two genes at any locus of a chromosome pair). A *DNA profile* consists of measurements on the genotype at a number of *forensic markers*, which are specially selected *loci* on different chromosomes. Each *genotype* consists of an unordered pair of genes, one inherited from the father and one from the mother (though one cannot distinguish which is which). When both alleles are identical the actor is *homozygous* at that marker, and only a single allele value is observed; otherwise the actor is *heterozygous*.

Current technology uses around 8–20 *short tandem repeat* (STR) markers. At each marker, each gene has a finite number (up to around 20) of possible values, or *alleles*, generally positive integers. For example, an allele value of 5 indicates that a certain word (e.g., *CAGGTG*) in the four letter alphabet is repeated exactly 5 times in the DNA sequence at that locus.

In statistical terms, a gene is represented by a random variable, whose realized value is an allele.

In a particular forensic context, we will refer to the various human individuals involved in the case as "actors." An actor's *DNA profile* comprises a collection of genotypes, one for each marker.

Assuming *Mendelian segregation*, at each marker a parent passes a copy of just one of his or her two genes, randomly chosen, to the child, independently of the other parent and independently for each child. Databases have been gathered from which allele frequency distributions, for various populations, can be estimated for each forensic marker. Throughout this paper, our numerical examples use the allele frequencies reported in Butler et al. (2003).

The Hardy–Weinberg law states that the relative proportion of genotypes, with respect to a given locus, remains constant in a population so long as mating is random. If there is independence in the inheritance of genes at loci on the same chromosome pair, these loci are said to be unlinked. In standard forensic identification problems it is customary to assume the Hardy–Weinberg equilibrium, and that loci are unlinked, which corresponds to assuming independence within and across markers.

## 3. Methodology for assessing sensitivity.

3.1. *Bayesian networks.* A *Bayesian network* (BN) is a discrete multivariate probability model represented as a directed acyclic graph (DAG). Its formulation as a network provides a joint graphical and numerical representation allowing the application of fast general-purpose algorithms to compute inferences. In a Bayesian network, complex interrelationships are broken down into simple local dependencies, from which a full graphical representation can be built in a modular fashion. A genetic pedigree fits naturally into this framework: the family relationships constitute the local graphical modules, with the required conditional probability tables being simply specified, for example, by Mendel's laws of inheritance, or by the logical relationship between a genotype and its constituent genes.

We give a brief introduction to the basic elements of a Bayesian network; for further details see Cowell et al. (1999). The DAG $D$ underlying a Bayesian network represents qualitative relationships of dependence and independence between variables; $D$ consists of a set $V$ of nodes, and directed links, drawn as arrows. Each node $v \in V$ represents a random variable $X_v$. The set $pa(v)$ of parents of a node $v$ are those nodes in $D$ out of which arrows into $v$ originate. The quantitative structure of a Bayesian network is expressed in terms of a set of conditional probability distributions. The full joint probability density of $(X_v, v \in V)$ is defined by $p(x) = \prod_{v \in V} p(x_v | x_{pa(v)})$. Algorithms such as that by Lauritzen and Spiegelhalter (1988) transform the graph $D$ into a new representation called a junction tree of cliques which allows efficient computation of the conditional probability $p(x_v|x_A)$, for any $v \in V$, any set of nodes $A \subseteq V$ and any configuration $x_A$ of the nodes $X_A$. The nodes in A would typically be those at which we observe and input *evidence* $X_A = x_A$. A node $v$ at which the conditional distribution given the evidence is desired might be termed a *target node*.

3.2. *Marginal posteriors in a Bayesian network.* For forensic genetics it is useful to partition the set of variables in the joint probability model (which correspond to a set of nodes in the corresponding Bayesian network model) disjointly as

$$X = F \cup E \cup T \cup O, \tag{1}$$

corresponding to Founding genes, Evidence (that is, data), Targets and Others. We suppose that there is a single binary target, taking values $T = 1$ and 0, which correspond, for example, in criminal identification, to the true/false state of the hypothesis that the suspect left a trace at the scene of a crime.

In forensic genetics problems the weight of the evidence in favor of a hypothesis is generally expressed as a likelihood ratio. Our focus of attention in this paper is, therefore,

$$h(f) = \log \frac{P\{T = 1|E\}}{P\{T = 0|E\}},$$

the logarithm of the likelihood ratio (LR) for the hypothesis represented by $T$, given the data or evidence $E$. This will be expressed as a function of the joint distribution $f$ of the founders $F$, written as a vector indexed by the vector of values of the founding alleles: $P\{F = \mathbf{i}\} = f_{\mathbf{i}}$. Let $p_{t\mathbf{i}} = P\{T =$ $t, E|F = \mathbf{i}\}$ for $t = 0, 1$ and all $\mathbf{i}$, and write $p_t$ for the vector of $p_{t\mathbf{i}}$ for all $\mathbf{i}$, so that $P\{T = t, E\} = p_t^T f$. Then

$$\begin{aligned} h(f) &= \log \frac{P\{T = 1|E\}}{P\{T = 0|E\}} \\ &= \log \frac{P\{T = 1, E\}}{P\{T = 0, E\}} \\ &= \log \frac{\sum_{\mathbf{i}} P\{T = 1, E|F = \mathbf{i}\} f_{\mathbf{i}}}{\sum_{\mathbf{i}} P\{T = 0, E|F = \mathbf{i}\} f_{\mathbf{i}}} \\ &= \log \frac{p_1^T f}{p_0^T f}. \end{aligned}$$

The basis for our study is to evaluate variations in the value of $h(f)$ as $f$ varies from the default or *baseline* assumptions $f_0$ (typically homogeneous population, known frequencies, Hardy–Weinberg). In the following subsections we discuss several approaches to doing so.

3.3. *Representing dependence by structure.* The most straightforward approach to the numerical assessment of sensitivity of $h(f)$ to specific changes in $f$ is simply to set up and run a Bayesian network (BN) for a variety of alternative settings for $f$. This need not be too cumbersome for a small collection of alternative $f$s if the BN calculation can be conducted in a suitable programming environment (see Appendix A.3).

Many of the alternative $f$s that will be of interest, unlike the baseline $f_0$, will impose dependence among founding genes. This arises in the case of uncertainty in allele frequencies, for identity by descent, and often in the presence of subpopulation structure. Dependence can of course be handled within the discrete BN formalism, by elaborating the DAG of the model with additional parent–child connections between founding genes, as necessary. It is immediate to see how to do this for cases of subpopulation structure; methods for dealing with uncertain allele frequencies and identity by descent through model structure are deferred to Section 5.2 and Section 5.3 respectively.

3.4. *Multiple markers.*

3.4.1. *Marker data may not be conditionally independent.* Forensic genetics routinely uses from 8 up to 20 markers simultaneously, in order to increase the power of the inference. Thus, the evidence $E$ has a component $E_m$ for a number of markers $m = 1, 2, \ldots, N_m$. The standard assumption is that the $\{E_m\}$ are independent, given $T$, which arises by design, since markers are generally chosen from different chromosomes (and to be neutral in selection terms). In such a case, we have immediately that likelihood ratios for $T$ obey the "*product rule*":

$$\frac{P\{E|T=1\}}{P\{E|T=0\}} = \prod_{m=1}^{N_m} \left\{ \frac{P\{E_m|T=1\}}{P\{E_m|T=0\}} \right\}, \tag{2}$$

so that BN calculations can be run separately for each $m$, and trivially aggregated for the required combined inference.

However, if there are unobserved variables, other than $T$, common to all markers and correlated with them, then this conditional independence, and the product rule, will fail. As we will see, this applies to identity by descent, and to some cases of population substructure.

The most straightforward approach to dealing with the complication of multiple markers, when the product rule (2) fails, is to extend the model to handle all markers simultaneously. This is fairly routine if the structural approach of Section 3.3 is being used, given a suitable programming environment. However, the computational time and space requirements of a BN to handle all markers simultaneously typically grow rapidly with the number of markers, so it is of interest to seek alternative approaches.

3.4.2. *Computing across-marker inferences using within-marker BNs.* Consider the following joint probability model for marker data $E = \{E_m, m = 1, 2, \ldots, N_m\}$. There is a latent variable $R$ typically coding the relationship between the actors, and a target variable $T$ of interest. In terms of the general notation of Section 3.2, $R$ is part of $O$.

We assume

$$p(T, R, E) = p(T)p(R) \prod_{m=1}^{N_m} p(E_m|T, R), \tag{3}$$

that is, that markers are conditionally independent, given only the target node $T$ and the relationship variable $R$.

To calculate likelihood ratios between values of $T$, we need the marginal likelihoods $P(E|T)$, which can be expressed

$$\begin{aligned} p(E|T) &= \sum_R p(R) \prod_m p(E_m|T, R) \\ &= p(T)^{-N_m} \sum_R p(R) \prod_m p(E_m, T|R) \end{aligned}$$

since $T$ and $R$ are independent *a priori*. Probability propagation algorithms, such as those presented by Lauritzen and Spiegelhalter (1988) and Lauritzen (2003), when run on a Bayesian network with evidence $E_m$, for each marker $m$ and value of $R$ separately, deliver precisely $p(E_m, T|R)$, providing their output is left unnormalized.

The correct overall marginal likelihoods can thus be obtained simply by multiplying the BN output tables over markers and then averaging with respect to $P(R)$.

At the same time, the marker-specific likelihood ratios can be obtained from

$$p(E_m|T) = p(T)^{-1} \sum_R p(R) p(E_m, T|R)$$

and the (incorrect) answer obtained from the "product rule" is the result of multiplying these values together; it is clear that (2) does not hold. In brief, the product rule is in error by averaging the marginal likelihoods $p(E_m, T|R)$ over $R$ *before* multiplying over $m$.

This approach is available whenever (3) applies, whatever the interpretation of the latent variable $R$, and will be practical, providing $R$ does not take too many distinct values.

A more subtle variation can reduce the scale of the computation. Suppose there are within-marker latent variables $\pi = \{\pi_m, m = 1, 2, \ldots, N_m\}$ (in the case of identity by descent, these code the pattern of identity among genes for the respective markers), and suppose

$$p(T, R, \pi, E) = p(T)p(R) \prod_{m=1}^{N_m} \{p(\pi_m|R)p(E_m|T, \pi_m)\}. \tag{4}$$

Then the marginal likelihood can be manipulated as follows:

$$\begin{aligned} p(E|T) &= \sum_R p(R) \prod_m \left\{ \sum_{\pi_m} p(\pi_m|R) p(E_m|T, \pi_m) \right\} \\ &= \sum_R p(R) \prod_m \left\{ \sum_{\pi_m} p(\pi_m|R) p(E_m, T|\pi_m)/p(T|\pi_m) \right\} \\ &= p(T)^{-N_m} \sum_R p(R) \prod_m \left\{ \sum_{\pi_m} p(\pi_m|R) p(E_m, T|\pi_m) \right\}. \end{aligned}$$

This demonstrates that the required combined inference can also be obtained from within-marker BN calculations for each marker $m$ and each value of the latent variable $\pi_m$. Each BN in this case will be somewhat simpler since the global latent variable(s) $R$ are not involved.

The relative computational cost of the two alternative calculations depends on the numbers of distinct values taken by $R$ and by $\{\pi_m\}$.

3.5. *Constrained steepest descent (CSD).* A more analytic and potentially more general approach to deal with whole classes of alternative $f$ is to aim to bound differences $|h(f) - h(f_0)|$ in terms of $\|f - f_0\|$ and, in particular, study this for infinitesimal departures from $f_0$. In the absence of constraints on $f$, the direction in which $h$ varies most steeply is of course given by the gradient of $h$.

The gradient of the log LR $h$ at $f$ satisfies

$$((\nabla h)(f))_{\mathbf{i}} = \frac{p_{1\mathbf{i}}}{p_1^T f} - \frac{p_{0\mathbf{i}}}{p_0^T f},$$

that is,

$$(\nabla h)(f) = (p_1^T f)^{-1} p_1 - (p_0^T f)^{-1} p_0 = g,$$

say.

In practice, there will be constraints on $f$; in particular, it must be a probability distribution, necessitating bound constraints for nonnegativity and the equality constraint $f^T \mathbf{1} = 1$. We may also wish to impose symmetry constraints, arising from considerations of exchangeability between certain actors, or their genes, and so on. In this paper we consider only linear equality constraints on $f$, together with the ubiquitous nonnegativity bound constraints. Since $f$ is representing the (completely general) founding gene distribution as a vector of (joint) probability values, linear constraints on such vectors form a very general class of constraints—on the values of any moments, or probabilities, for example.

The constrained direction of steepest descent/ascent is the projection of the gradient vector of the objective function at the point in question onto the orthogonal complement of the constraints.

More explicitly, given a real function $h$ of a $n$-vector argument, the $n$-vector $\delta$ that maximizes $\lim_{\varepsilon \to 0} \varepsilon^{-1} |h(f_0 + \varepsilon\delta) - h(f_0)|$ subject to $\|\delta\| = 1$ and $X^T \delta = 0$ is given by $\delta = (I - H)g / \|(I - H)g\|$, where $g$ is the gradient of $h$ at $f_0$, $H = X(X^T X)^- X^T$ and $\| \cdot \|$ denotes euclidean norm. In the language of linear models, we regress $g$ on $X$ and scale the residuals to have norm 1.

One approach to reporting sensitivity of inferences to perturbations to $f_0$ of magnitude $\varepsilon$ is to deliver the maximum and minimum of $h(f)$ for $f$ lying on the line of constrained steepest descent, subject to $f_i \geq 0$, $\|f - f_0\| < \varepsilon$. These cannot strictly be interpreted as bounds, since they are based on a linearization of $h(f)$ and because the nonnegativity constraint may by chance bite particularly severely in the constrained steepest descent direction.

It may also be of interest to weight the coordinate directions unequally, replacing the spherical neighborhood of $f_0$ implicit in the derivation above by an ellipsoidal one. This would allow, for example, approximating *relative* departures from $f_0$ instead of absolute ones. Given a symmetric positive definite matrix $W$, we then seek the $n$-vector $\delta$ to maximize $\lim_{\varepsilon \to 0} \varepsilon^{-1} |h(f_0 + \varepsilon W\delta) - h(f_0)|$ subject to $\|\delta\| = 1$ and $X^T W \delta = 0$. The optimizing direction $\delta$ is $W(I - H_W)g / \|W(I - H_W)g\|$, where $H_W = X(X^T W^2 X)^- X^T W^2$.

3.6. *Linear fractional programming (LFP).* A second analytic method exploits the linear fractional form of $\exp(h(f)) = p_1^T f / p_0^T f$ as a function of $f$. Linear fractional programming concerns the problem of minimizing a function of the form

$$\frac{\alpha_0 + \boldsymbol{\alpha}^T \mathbf{x}}{\beta_0 + \boldsymbol{\beta}^T \mathbf{x}}$$

over nonnegative variables $\mathbf{x} = (x_1, x_2, \ldots, x_n)$ subject to linear equality or inequality constraints. There are various approaches [Bajalinov (2003)], but the most straightforward reduces this to a standard linear programming problem. Expositions of this approach are either sketchy or rather inaccessible [Vajda (1975); Gass (1969); Charnes and Cooper (1962)], so we give the basic details here.

Suppose the linear constraints are of the form

$$\sum_{j=1}^{n} a_{ij} x_j \le b_i \qquad \text{for } i = 1, 2, \ldots, c.$$

Other equality or inequality constraints, in any combination, are treated similarly. We suppose that the set of feasible $\mathbf{x}$ is bounded. Then it is readily shown that the optimization can be performed by running two artificial linear programs, each minimizing

$$\alpha_0 y_0 + \boldsymbol{\alpha}^T \mathbf{y}$$

over the $(n+1)$ variables $(y_0, \mathbf{y} = (y_1, y_2, \ldots, y_n))$ subject to

$$\sum_{j=1}^{n} a_{ij} y_j - b_i y_0 \le 0 \qquad \text{for } i = 1, 2, \ldots, c.$$

$$\beta_0 y_0 + \boldsymbol{\beta}^T \mathbf{y} = \delta \quad \text{and} \quad y_0, y_1, \ldots, y_n \ge 0,$$

where $\delta = +1$ in one problem and $-1$ in the other. The solution $(y_0^\star, \mathbf{y}^\star)$ for whichever of these problems gives the smaller minimum is used (often, in fact, only one of the two problems has feasible solutions) and then

$$x_j^\star = \frac{y_j^\star}{y_0^\star}$$

gives the optimum for the original problem.

Returning to our problem of interest, linear fractional programming allows us to find the minimum and maximum of the $\log \mathrm{LR} = h(f)$, subject to an arbitrary set of linear constraints $X^T f = X^T f_0$, as in Section 3.5, and linear bounds on the difference between $f$ and $f_0$, for example, $\max_{\mathbf{i}} |(f - f_0)_{\mathbf{i}}| \le \varepsilon$ (i.e. that the total variation norm is less than $\varepsilon$).

In this setting the calculations can be simplified, since the coefficients in the linear combinations of $f_{\mathbf{i}}$ in the numerator and denominator of the posterior odds are nonnegative. It then turns out that the LP problem with $\delta = -1$ is never feasible, so only the $\delta = +1$ computation need be done.

As in Section 3.5, we can readily weight the coordinate directions unequally, and amend this to seek the extremes of the log LR within the hyper-rectangle $\max_{\mathbf{i}} |(f - f_0)_{\mathbf{i}}/w_{\mathbf{i}}| \le \varepsilon$—the resulting bounds are still linear.

For both this approach and that of constrained steepest descent, Section 3.5, the dimension of the free variable $f$ in the optimization scales exponentially with number of markers, so the only realistic possibility of using these bounds numerically for multiple markers, when markers are dependent, would involve exploiting the identities in Section 3.4.2.

**4. Example settings.** We consider four examples, two from criminal law, a case of simple criminal identification and a DNA mixture, and two from paternity cases, simple paternity and disputed sibship. The data for the DNA mixture and the paternity cases were based on real forensic casework. In all the examples given there are two competing hypotheses $H_0$ and $H_1$ and the strength of the evidence in favor of $H_0$ is given by the likelihood ratio

$$\mathrm{LR} = \frac{P(E|H_0)}{P(E|H_1)},$$

where the evidence $E$ consists of measurements on a set of DNA markers. The hypotheses $H_0$ and $H_1$ correspond to the two values of the Boolean target $T$ of Section 3.2.

We introduce the four examples in turn, with sample data, using a mix of algebraic and graphical formulations of the probability models we need. Only the first example is covered in full detail; for the other examples, in order to save space, we concentrate only on the additional features each introduces.

4.1. *Criminal identification.* A pictorial representation of a simple criminal identification case, with a single charge against a single suspect, together with its expanded version, is shown in Figure 1. The evidence might be that the suspect's DNA profile matches the one found at the crime scene. Suppose we are interested in testing the prosecution hypotheses $H_0$: *the crime trace belongs to the suspect* `s` (loosely, "the suspect is guilty"); versus the defense hypothesis $H_1$: *the crime trace belongs to another actor* `as` *randomly drawn from the population.* Representation of such problems as Bayesian networks was introduced by Dawid et al. (2002), and as object-oriented Bayesian networks by Dawid, Mortera and Vicard (2007).

The target $\mathtt{T}$ of the inference is the Boolean variable $\mathtt{S\ guilty?}$, whose values *true* and *false* correspond to $H_0$ and $H_1$ respectively.

The two actors, $\mathtt{s}$ and $\mathtt{as}$, are each fully described by three variables for each marker $m$, the paternal gene, maternal gene and genotype, denoted, for example, by $(\mathtt{spg}_m, \mathtt{smg}_m, \mathtt{sgt}_m)$. For each marker and each actor, the genotype is determined as the logical combination of the corresponding genes, for example, $p(\mathtt{asgt}_m = \{8, 11\} | \mathtt{aspg}_m = 8, \mathtt{asmg}_m = 11) = 1$. All genes for both actors are initially assumed drawn i.i.d. from prescribed allele frequencies for the corresponding marker.

For each marker, $\mathtt{trace}_m$ represents the crime scene trace for that marker, and is modeled as identical to $\mathtt{sgt}_m$ or $\mathtt{asgt}_m$ according to whether $\mathtt{S\ guilty?}$ is *true* or *false*, respectively. For example, $p(\mathtt{trace}_m = \mathtt{sgt}_m | \mathtt{sgt}_m, \mathtt{asgt}_m, \mathtt{S\ guilty?} = \mathit{true}) = 1$. Our task is to compute the likelihood ratio for $\mathtt{S\ guilty?}$ corresponding to the observed evidence that $\mathtt{trace}_m$ and $\mathtt{sgt}_m$ coincide, with values given in Table 1.

We can write the joint distribution of all variables as

$$
\begin{aligned}
&p(\mathtt{S\ guilty?}) \prod_m [p(\mathtt{spg}_m) p(\mathtt{smg}_m) p(\mathtt{aspg}_m) p(\mathtt{asmg}_m)] \\
&\quad \times \prod_m [p(\mathtt{sgt}_m | \mathtt{spg}_m, \mathtt{smg}_m) p(\mathtt{asgt}_m | \mathtt{aspg}_m, \mathtt{asmg}_m) \\
&\qquad \times p(\mathtt{trace}_m | \mathtt{sgt}_m, \mathtt{asgt}_m, \mathtt{S\ guilty?})],
\end{aligned} \tag{5}
$$

the conditional independence structure of which is represented at a block level by the DAG in Figure 1. In the expanded version of the network in Figure 1, the variables within each block are visible, although the inner DAG structure is hidden. It shows the correspondence to the factors in (5), and the blocks are annotated according to partition (1).

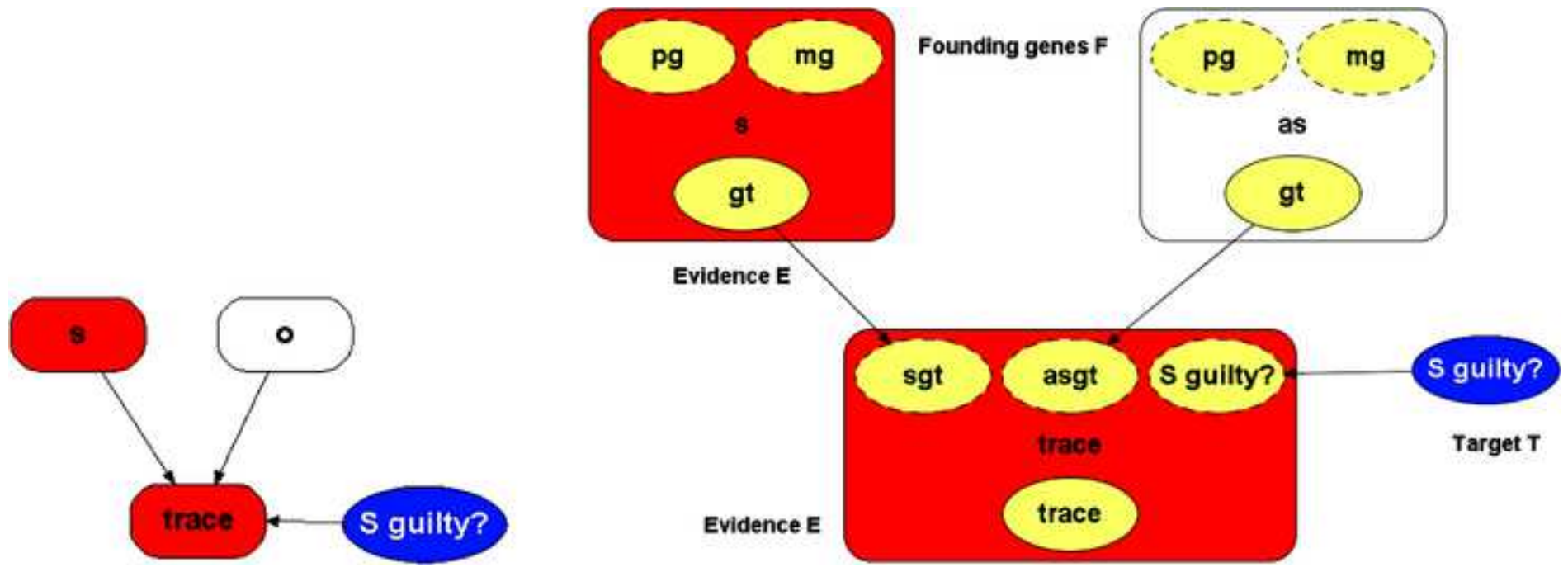

Fig. 1. *(Left) Block-level network for criminal identification, showing actors* $\mathtt{s}$ *and* $\mathtt{as}$*, crime trace and target. (Right) Expanded network revealing variables within each block node.*

TABLE 1
*DNA profiles of crime scene trace and suspect*

| **Marker** | **D13** | **D3** | **D5** | **D7** | **FGA** | **THO1** | **TPOX** | **VWA** |
|---|---|---|---|---|---|---|---|---|
| `sgt` & `trace` | 9 14 | 11 17 | 9 11 | 10 10 | 21 22 | 7 7 | 10 11 | 18 18 |

TABLE 2
*Data showing mixture composition, suspect's and victim's genotypes*

| **Marker** | `mix` | `sgt` | `vgt` |
|---|---|---|---|
| D13 | 8 11 | 8 8 | 8 11 |
| D3 | 16 18 | 18 18 | 16 16 |
| D5 | 12 13 | 12 13 | 12 12 |
| D7 | 8 10 11 | 8 10 | 8 11 |
| FGA | 22 24 25 26 | 22 26 | 24 25 |
| THO1 | 6 7 | 6 7 | 6 7 |
| TPOX | 8 11 | 8 8 | 8 11 |
| VWA | 17 18 | 17 17 | 17 18 |

In this paper we examine departures from the baseline assumptions of independence of all the variables $\{\mathtt{spg}_m, \mathtt{smg}_m, \mathtt{aspg}_m, \mathtt{asmg}_m, m = 1, 2, \ldots, N_m\}$, so the factor $\prod_m [p(\mathtt{spg}_m)p(\mathtt{smg}_m)p(\mathtt{aspg}_m)p(\mathtt{asmg}_m)]$ in (5), corresponding to baseline $f_0$, will in general be replaced by a different but appropriate joint distribution.

4.2. *Mixed trace.* When several actors may have contributed to a DNA sample left at a crime scene we encounter the problem of *mixed traces*. Here we consider a mixed trace, based on a real murder case which took place in Firenze. The mixture was assumed to be from two actors and we wish to test the hypothesis $H_0$: `s&v` that the suspect `s` and the victim `v` contributed to the mixture, as compared to the hypothesis $H_1$: `as&v` that an unknown actor in the population, `as`, and the victim contributed to the mixture. One might alternatively consider an additional unknown actor `av` instead of the victim, in which case the hypotheses are $H_0$: `s&av` and $H_1$: `as&av`. For a general description of the problem of DNA mixtures we refer to Mortera, Dawid and Lauritzen (2003). The evidence $E$ consists of the suspect's and the victim's genotypes and the DNA mixture composition as shown in Table 2. The presence of three and four alleles for markers D7 and FGA, respectively, are a clear indication that the trace is a mixture from more than one contributor.

Figure 2 shows the top-level network which can be used for analyzing a mixture with two contributors, `p1` and `p2`. Nodes `s`, `v`, `as` and `av` represent the suspect, the victim and two unknown actors. Only the genotypes, `sgt` etc., for the four actors contribute to the rest of the model specification. Boolean node `p1=s?` represents the hypothesis that contributor `p1` is the suspect. The variable $\texttt{p1gt}_m$ selects between $\texttt{sgt}_m$ and $\texttt{asgt}_m$ according to the true/false state of the Boolean variable `p1=s?`, in the same way as `S guilty?` is used to switch between $\texttt{sgt}_m$ and $\texttt{asgt}_m$ in the previous section. A similar relationship holds between the variables $\texttt{p2gt}_m$ $\texttt{vgt}_m$, $\texttt{avgt}_m$ and `p1=v?`. The `target` node is the logical combination of the two Boolean nodes `p1=s?` and `p2=v?` and represents the four different hypotheses described above.

In the *baseline model* for this example, the founding actors `s`, `v`, `as` and `av` each have paternal and maternal genes drawn for each marker independently from the gene pool for the appropriate population. This completes the joint distribution for all variables in the model which could be written out in expanded form as in equation (5), but in this and subsequent examples, we suppress this representation to save space.

Genotype information on the suspect and/or the victim is entered by fixing the values of $\texttt{sgt}_m$ and $\texttt{vgt}_m$. The variable $\texttt{mix}_m$ represents the mixed trace given by all possible combinations of alleles from contributors `p1` and `p2`, and information on the alleles seen in the mixture is entered there. All the evidence can be propagated by the Bayesian network calculations to find the required marginal posterior distribution for `target`.

4.3. *Simple paternity testing.* In a simple disputed paternity case, shown in Figure 3, we have an alleged family triplet formed by a disputed child `c`, its undisputed mother `m` and the putative father `pf`. DNA profiles are obtained from `c`, `m` and `pf`. On the basis of this evidence $E$, we wish to assess the likelihood ratio for the hypothesis of paternity, $H_0$: `tf=pf?` $=$ *true*, the true father tf is the putative father; as against that of nonpaternity $H_1$: tf=pf? $=$ *false*, the true father is an alternative actor af, randomly drawn from the population. Thus, tf=pf? switches deterministically between two alternatives, as seen in the previous examples; the only difference here is that this switching is now at the level of paternal and maternal genes, not the genotype. The corresponding factors in the joint probability model are therefore

$$\prod_m [p(\texttt{tfmg}_m|\texttt{pfmg}_m, \texttt{afmg}_m, \texttt{tf=pf?})p(\texttt{tfpg}_m|\texttt{pfpg}_m, \texttt{afpg}_m, \texttt{tf=pf?})].$$

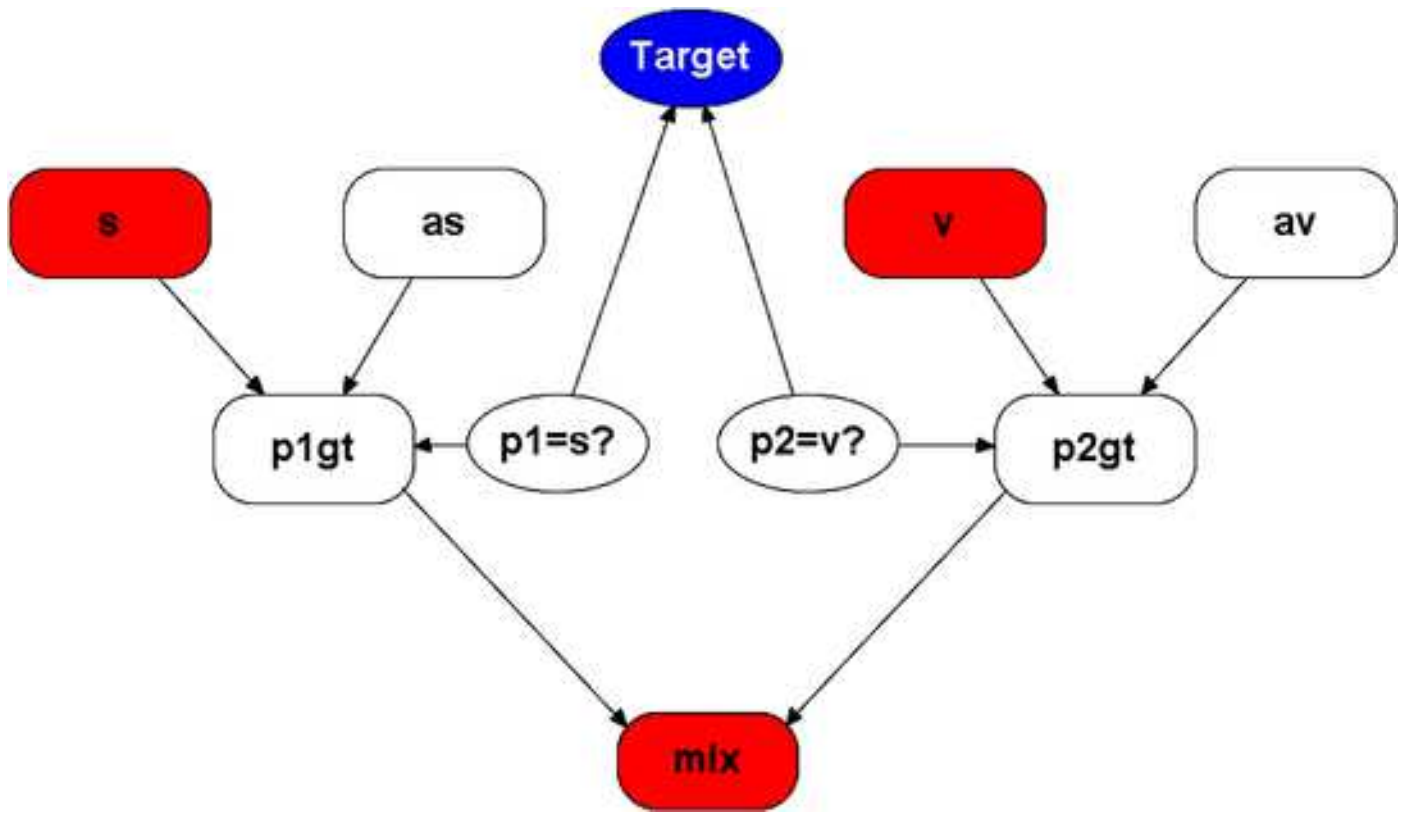

FIG. 2. *Network for a DNA mixture from two contributors.*

Table 3
*Paternity testing data showing genotypes of mother, child and putative father*

| **Marker** | mgt | cgt | pfgt |
|---|---|---|---|
| D13 | 10 13 | 13 13 | 11 13 |
| D3 | 16 17 | 16 17 | 17 18 |
| D5 | 11 12 | 11 11 | 11 11 |
| D7 | 10 12 | 10 11 | 11 12 |
| FGA | 23 23 | 21 23 | 21 23 |
| THO1 | 6 6 | 6 7 | 7 7 |
| TPOX | 8 11 | 8 11 | 11 11 |
| VWA | 18 18 | 18 18 | 17 18 |

The genes are needed since in Figure 3 the representation of c as a child of m and tf signifies the independent random draws of the child's genes from those of its parents according to Mendel's law, independently for each marker.

The baseline assumptions in this case will be that the paternal and maternal genes at each marker for each of m, af and pf are drawn independently from the relevant populations.

Table 3 gives the paternity testing evidence that we analyze in Section 6.

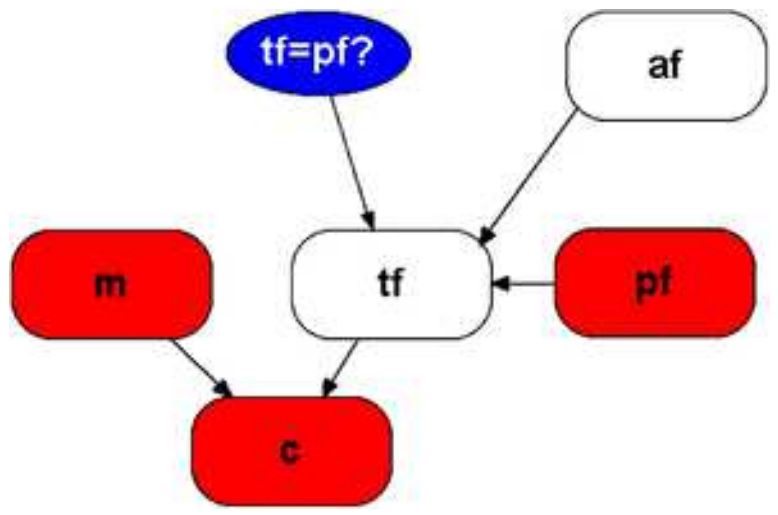

Fig. 3. *Pedigree for simple disputed paternity.*

TABLE 4
*Disputed sibship data*

| **Marker** | `m1gt` | `c1gt` | `m2gt` | `c21gt` | `c22gt` |
|---|---|---|---|---|---|
| D13 | 10 13 | 13 13 | 12 12 | 11 12 | 12 13 |
| D3 | 16 17 | 16 17 | 15 18 | 15 18 | 15 18 |
| D5 | 11 12 | 11 11 | 11 11 | 11 11 | 11 11 |
| D7 | 10 12 | 10 11 | 10 10 | 10 10 | 10 10 |
| FGA | 23 23 | 21 23 | 24 25 | 20 25 | 20 24 |
| THO1 | 6 6 | 6 7 | 9 9.3 | 7 9 | 7 9.3 |
| TPOX | 8 11 | 8 11 | 10 11 | 10 10 | 10 10 |
| VWA | 18 18 | 18 18 | 17 20 | 17 18 | 18 20 |

4.4. *Disputed sibship.* The pedigree in Figure 4 represents a real case of disputed inheritance, essentially a more complicated paternity dispute. We have an undisputed family with two children `m2`, `tf2`, `c21` and `c22` and it is questioned whether the deceased father `tf2` is also the true father `tf1` of a child `c1` by another mother `m1`. The target hypothesis of interest is represented by the Boolean node `tf1=tf2?` which embodies the two hypotheses that `tf2` is the father of `c1` or not, according as its value is *true* or *false*.

In this example, the baseline assumptions are that the paternal and maternal genes at each marker for each of `m1`, `m2`, `af` and `tf2` are drawn independently from the relevant populations. The complete joint distribution, following the schematic structure of Figure 4, is thus a product of terms of the kind met in earlier examples: factors for these founding genes, for the Mendelian inheritance of genes for the three children, and selection (of both genes for all markers) between `af` and `tf2` according to the value of `tf1=tf2?`, similarly to the selection in Section 4.3.

The data available are given in Table 4 and comprise only the genotypes of `m2`, `c21`, `c22`, `m1` and `c1`.

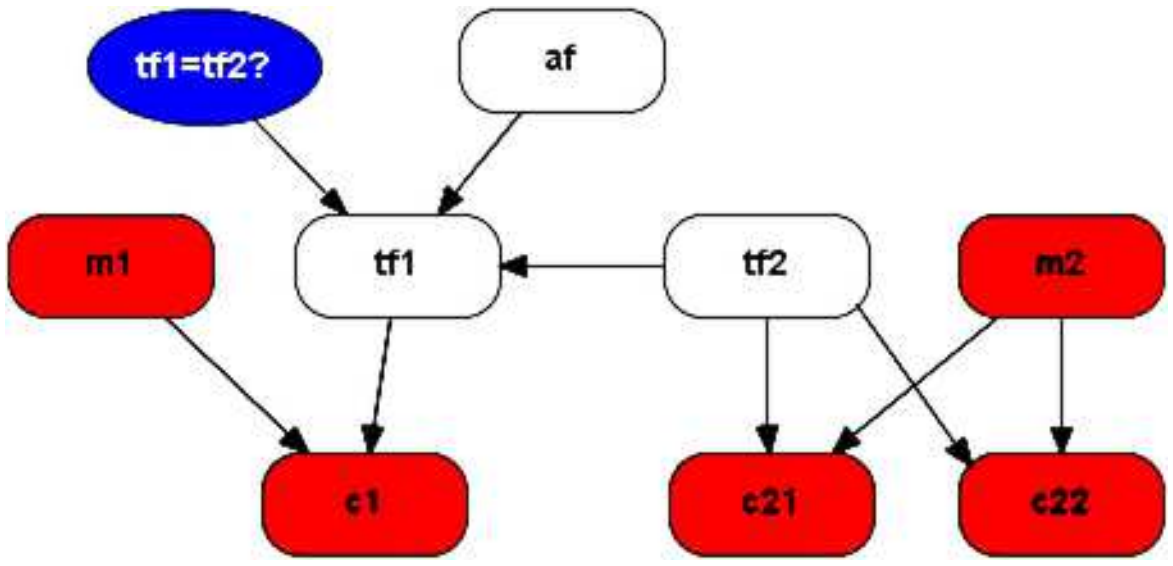

FIG. 4. *Pedigree for disputed sibship.*

**5. Variations on baseline assumptions.** We now describe the different scenarios—uncertainty in allele frequencies, identity by descent and population substructure—to be considered as departures from the baseline assumptions, in each of the four examples of the previous section. In each example the baseline is that, for each marker, all founding genes are drawn independently from defined "gene pools" with specified allele frequencies. Our variant scenarios all replace this baseline assumption by a different joint distribution for these founding genes, for which we propose appropriate probability models, again using a mix of algebraic and graphical formulations. For definiteness we will concentrate in the following discussion on the implications for simple criminal identification (Section 4.1). All of the other examples can be handled similarly.

5.1. *Coancestry coefficients.* The standard approach to allowing for departures from the baseline assumptions in genetic calculations is to capture the "ambient" degree of relatedness in a population by means of a single scalar parameter $\theta$, which we call here the coancestry coefficient. The concept is found in many different guises in the literature, reflecting the diversity of causes for dependence between genes drawn randomly from a gene pool, and the multiplicity of models for this dependence. Informally, $\theta$ is the "proportion of alleles that share a common ancestor in the same subpopulation" [Balding and Nichols (1994)]. It can be identified with Wright's measure of interpopulation variation $F_{ST}$ [Wright (1940, 1951)].

For a more explicit definition, suppose that $n$ genes have been drawn at random, of which $m$ are allele $a$, then the probability that the next gene is also allele $a$ is

$$\frac{m\theta + (1-\theta)\rho(a)}{1+(n-1)\theta} = \frac{m+\alpha\rho(a)}{n+\alpha}, \tag{6}$$

where $\rho(a)$ is the marginal probability of drawing the allele $a$, and $\alpha = (1-\theta)/\theta$. Used recursively, this relation determines the joint distribution of any finite set of genes $g_1, g_2, \ldots, g_n$. This can be written

$$\frac{\Gamma(\alpha)}{\Gamma(n+\alpha)} \prod_{i=1}^{n} \left\{ \sum_{j<i} \delta(g_j, g_i) + \alpha\rho(g_i) \right\}, \tag{7}$$

where $\delta(\cdot,\cdot)$ is the Kronecker delta, and $\rho(\cdot)$ is the marginal allele distribution. This joint distribution represents one of the variations $f$ to the baseline distribution $f_0$. The informal definition of $\theta$ mentioned above arises since for two genes we can easily derive the joint distribution $p(g_1, g_2) = (1-\theta)\rho(g_1)\rho(g_2) + \theta\delta(g_1, g_2)\rho(g_1)$.

We can interpret (6) as saying that the next gene is with probability $1/(n+\alpha)$ a copy of each of the preceding $n$ genes (whatever the patterns of equality among them), and with probability $\alpha/(n+\alpha)$ an independent random draw from the gene pool. We recognize this as the Pólya urn scheme [Blackwell and MacQueen (1973)] corresponding to a Dirichlet process model [Ferguson (1973)], albeit one with, unusually, a discrete base measure. A particular consequence is that the joint distribution (7) determined by (6) is exchangeable, so that the order in which genes are introduced does not matter.

The value $\theta = 0$ ($\alpha \to \infty$) corresponds to independent sampling from $\{\rho(\cdot)\}$, and positive values to positive dependence due to population coancestry. The model has been used ubiquitously to adjust for coancestry, whether due to identity by descent [e.g., Balding and Nichols (1995) and Ayres and Balding (2005)], population structure [e.g. Fung and Hu (2004)], or for other reasons.

The strength of this approach to dependence lies in the convenience of capturing complex patterns of dependence with multiple causes in a single number, the weakness is that it can be at best a crude approximation to suppose that the dependence among the founding genes in a particular setting follows such a simple process as (6).

5.2. *Uncertain allele frequencies (UAF).* In reality, the allele frequencies assumed when conducting probabilistic forensic inference against an assumed background population are not fixed probabilities, but empirical frequencies in a database. An imperfect idealization is to regard these databases as independently drawn random samples from corresponding populations. Assuming a Dirichlet $(\delta(1), \delta(2), \ldots, \delta(k))$ prior and multinomial sampling with sample size $n$, the posterior distribution of a set of probabilities $\mathbf{r} = (r(1), r(2), \ldots, r(k))$ is Dirichlet $(M\rho(1), M\rho(2), \ldots, M\rho(k))$, where $M = n + \sum_i \delta_i$, and $\boldsymbol{\rho} = (\rho(1), \rho(2), \ldots, \rho(k))$ are the posterior means.

Our model for uncertain allele frequencies is that the founding genes (`apg`, `amg`, `bpg` and `bmg`, for example) are drawn i.i.d. from the distribution $\mathbf{r}$ across alleles, which in turn has the above Dirichlet distribution in which $\boldsymbol{\rho}$ are the database allele frequencies. The variation in $\mathbf{r}$ induces dependence among `apg`, `amg`, `bpg` and `bmg`, but in contrast to the case of identity by descent (illustrated in Section 5.3), there is still independence across markers.

In general, drawing the founding genes $g_1, g_2, \ldots, g_n$ conditionally independently from $\mathbf{r}$, where $\mathbf{r}$ is in turn drawn from a Dirichlet prior, corresponds exactly to the standard set-up for a Dirichlet process model, specifically, that defined by (7), with $\alpha = M$. The Pólya urn scheme for this model can be stated explicitly as follows. The first gene $g_1$ is drawn from$\boldsymbol{\rho}$. Then with probability $1/(M + 1)$, $g_2$ is set equal to $g_1$, and otherwise is a fresh random draw from $\boldsymbol{\rho}$. In general, $g_n$ is equal to each of $g_1, g_2, \ldots, g_{n-1}$ with probability $1/(M + n - 1)$ each, and with the remaining probability $M/(M + n - 1)$ is an independent random draw from $\boldsymbol{\rho}$. This is the same mechanism as described in the previous subsection, so perhaps surprisingly the model (7) is a more appropriate description of uncertainty in allele frequencies than identity by descent.

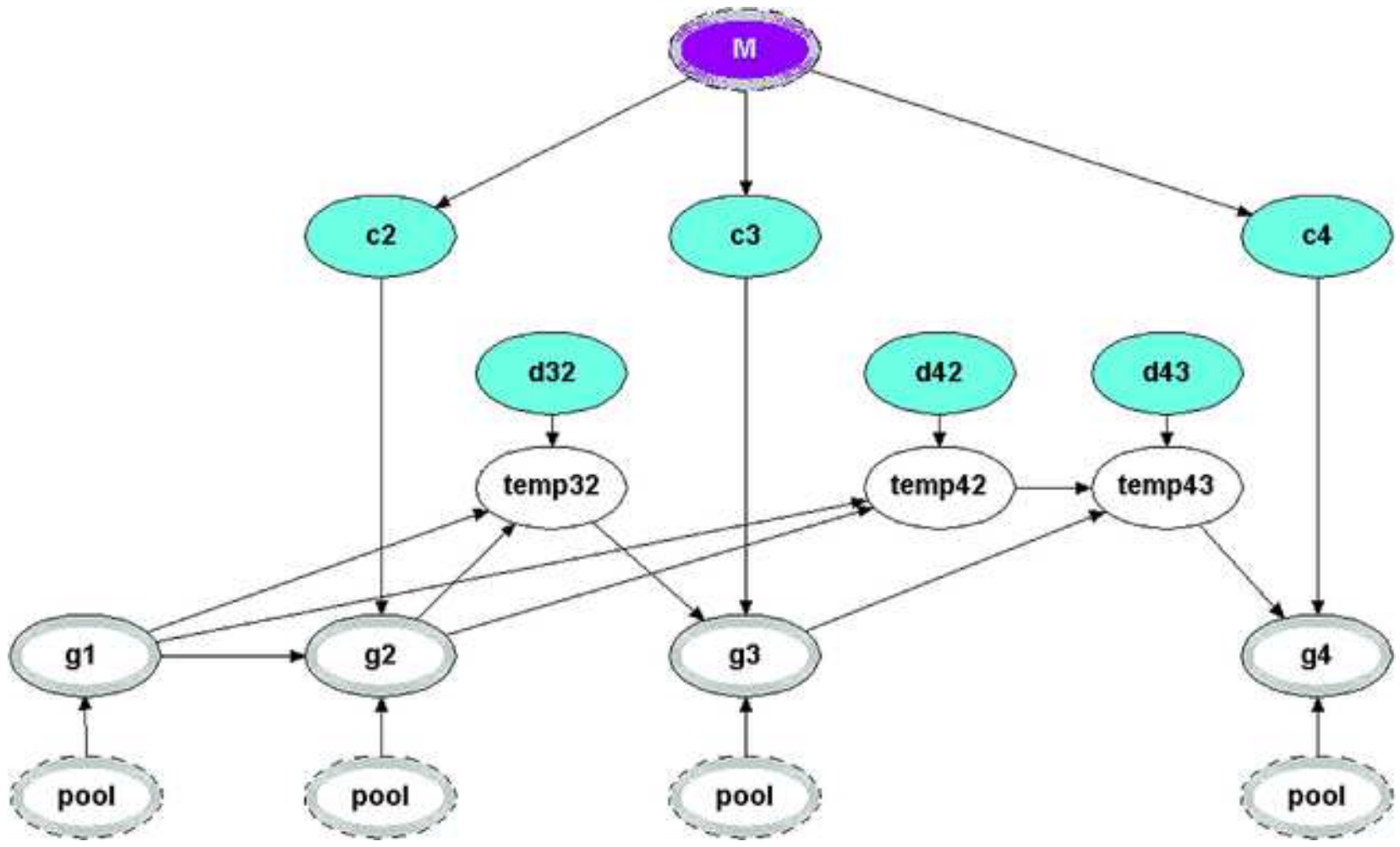

FIG. 5. *Network for Pólya urn scheme.*

The urn scheme is amenable to representation as a Bayesian network, with founder nodes that include the independent random draws from the gene pool, and terminal nodes the required $g_1, g_2, \ldots,$ with intermediate nodes carrying out the required switching. This can be set up in various ways; in the interests of computational time and space in probability propagation, it is generally best to organize the net so that all choices are binary. This procedure of "divorcing" creates smaller clique tables in the Bayesian network [Jensen (1966)]. The result is represented by a network whose DAG is shown in Figure 5 and which is set out in pseudo-code in Appendix A.1.

5.3. *Identity by descent (IBD).* When two actors, say, `a` and `b`, in a case are related, it is no longer correct to regard their genes `apg`, `amg`, `bpg` and `bmg` as random variables independent *a priori* (and independent across markers), since Mendelian inheritance from their common ancestor(s) induces dependence between their genes. This is called identity by descent [Cotterman (1974); Thompson (1974)]. Wright (1940) devised an "island model" for a compound population, in which a finite "island" population is regularly refreshed with immigration from an infinite "mainland" population with constant allele frequencies. The actors are drawn randomly from the island population, and the immigration bottleneck induces dependence among their genes that can be taken to describe identity by descent. Wright obtained a distribution equivalent to (6) for the dependence between genes. The same result has been derived for a more flexible birth–death–immigration process by Rannala (1996).

The standard practice of adjusting the formulae for the required likelihood ratios using the corresponding value of $\theta = F_{ST}$ as in Balding and Nichols (1995) assumes no specific relationship between the actors, so may give a poor approximation to the truth in situations where such relationships (or their probabilities) can be assumed. Further, the standard approach ignores the fact that relatedness, where the relationship is uncertain, induces dependence between markers, as mentioned in Section 3.4.1. Here we set up a probabilistic formulation that does capture all the dependencies, if the analyst is prepared to model the specific relationships between the actors probabilistically.

For any specified form of relationship, it is straightforward to use Mendel's law to derive the correct joint distribution of `apg`, `amg`, `bpg` and `bmg`, especially if the relationship is fairly close. For example, if `a` is father to `b`, then we find that `apg`, `amg` and `bmg` are independent draws from the gene pool, while, given these, `bpg` is equal to `apg` or `amg` with probabilities 1/2 each. As a second example, if `a` and `b` are siblings, then there are four possible patterns of identity among `apg`, `amg`, `bpg` and `bmg`, delivered with equal probability: there may be no IBD, or `bpg = apg`, or `bmg = amg`, or both of these, where in each case those variables not appearing to the left of a `=` sign are independent draws from the gene pool. A table of these patterns, giving further examples, can be found in Appendix A.2.

When the actual relationship between the actors is not known but we are prepared to assign it a prior probability distribution, then at each marker, the joint distribution of the founding genes is the corresponding weighted average. However, with multiple markers, the story is more complicated. We can build a full probability model for IBD using the formulation in the second half of Section 3.4.1. The variable $R$ signifies the relationship (e.g., `a` is father to `b`) and the variable $\pi_m$ the pattern of identity (e.g., `bpg = apg`, others different), for marker $m$. These patterns of dependence are very naturally expressed by elaborating the graphical model with these additional variables, and generating the founding genes from these nodes and independent draws from the gene pool. This is visualized in Figure 6. Note that, conditional on the relationship between the actors, the pattern of IBD among the actors' genes is independent from marker to marker.

The idea of using a latent "pattern" variable to structure the dependence among founding genes is very general, and should be useful in modeling departures from baseline other than those considered here. It could have been used in place of the Pólya urn scheme for UAF in Section 5.2, but at some cost in efficiency.

5.4. *Population heterogeneity (HET).* Population heterogeneity raises two kinds of issues in the modeling. First, since unobserved actors are assumed to have genes drawn from a population, results can depend on which population (and corresponding allele frequency database) is used. Second, when there is uncertainty about which population is relevant, this can induce dependence between actors, observed or not. Additionally, when uncertainty about subpopulation relates to untyped actors, dependence between markers is induced.

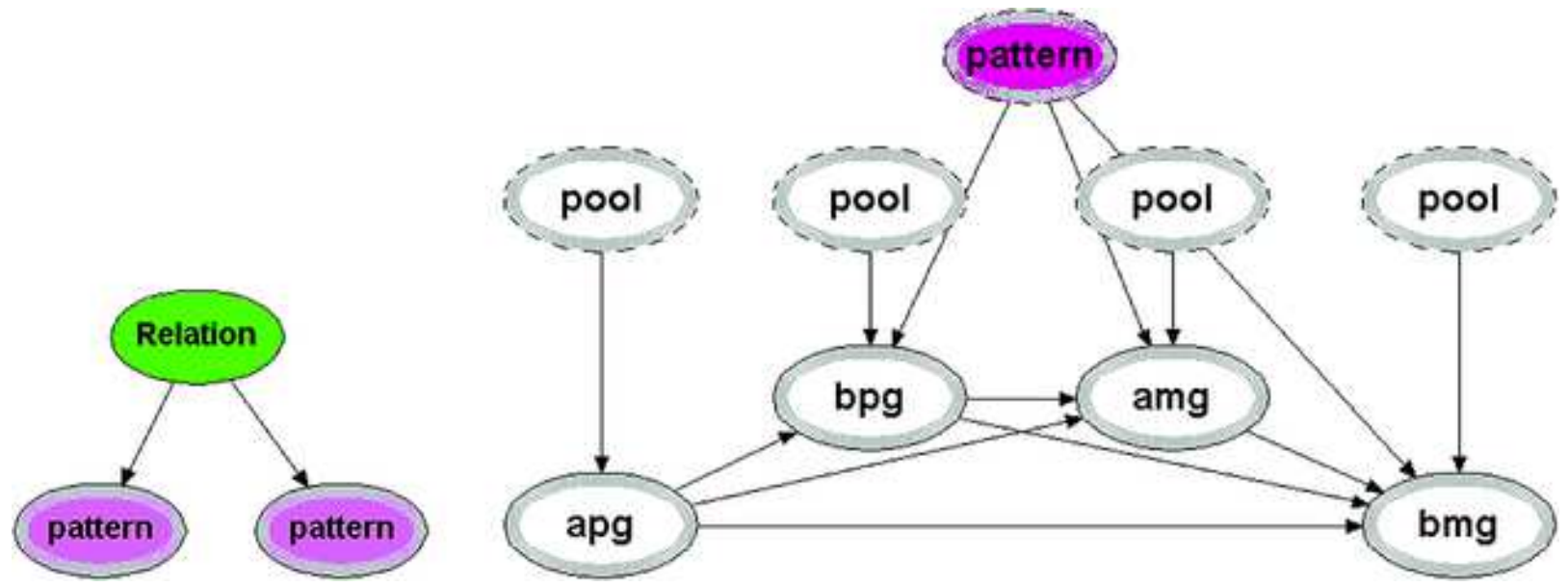

Fig. 6. *Networks representing (left) the dependence of pattern of IBD $\pi_m$ on relation $R$, for two markers and (right) dependence of paternal and maternal genes at marker $m$ for two individuals on pattern $\pi_m$.*

The sensitivity of the resulting inferences to population structure, based on a synthetic population that is a mixture of Afro-Caribbean, Hispanic and Caucasian subpopulations, is presented for various examples in Section 6. Such problems are easily set up as Bayesian networks with the structure shown in Figure 7, where `S` is a variable identifying the subpopulation, which may be dependent (perhaps identical) or independent between actors depending on the scenario of interest. Crucially, for each actor, `S` is the same for both genes for all markers, so that mixing across subpopulations is not the same as averaging the allele frequencies and assuming an undivided subpopulation. This observation may have wider implications; since all real populations are to a degree heterogeneous, some dependence between markers will be ubiquitous.

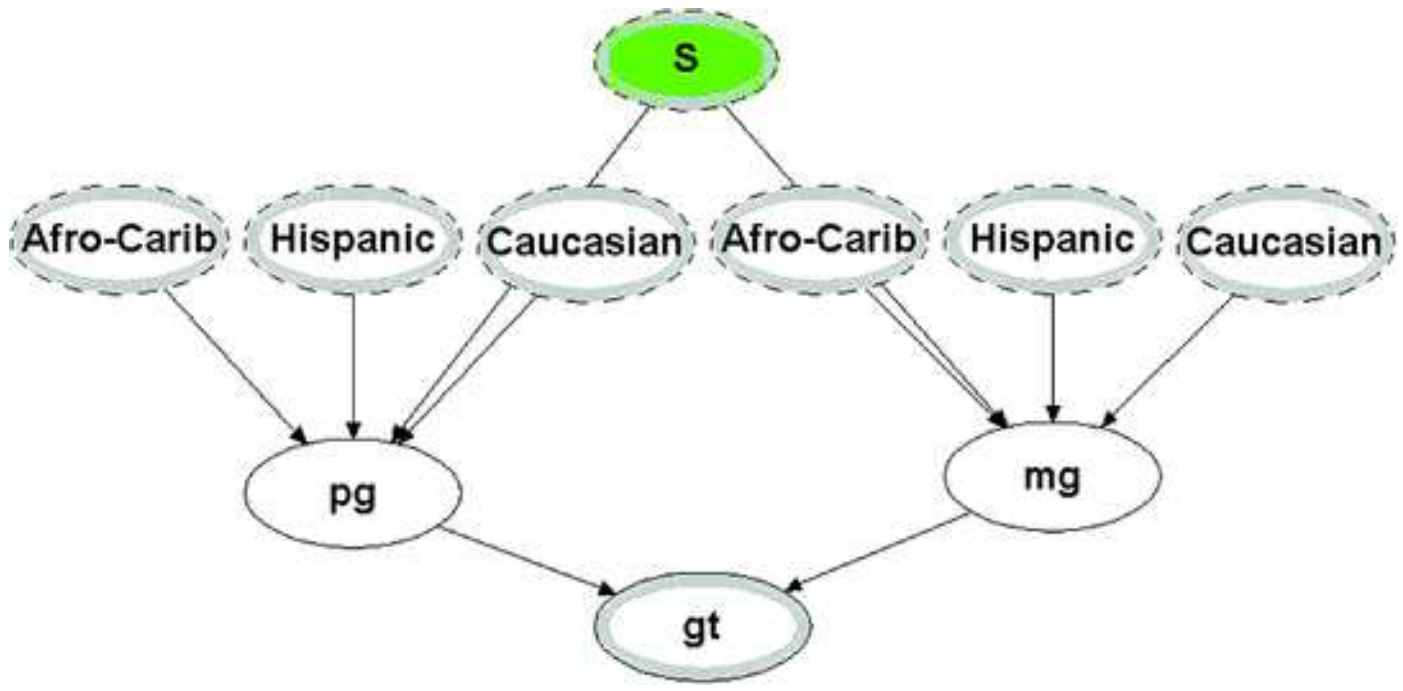

Fig. 7. *Network for genotype allowing for subpopulation effect; note that conditional on subpopulation $S$, every gene at every marker is drawn independently from the appropriate subpopulation gene pool.*

5.5. *Combinations of scenarios.* Bayesian networks are intrinsically modular, and so are our scenarios, so it should be no surprise that the scenarios can be combined. This raises further challenges to the $\theta/F_{ST}$ approach (6) to coancestry. In particular, it is important to note that the output from two modules of this form "cascaded" together does not follow this distribution.

To see this, recall the form of the partition distribution induced by the Dirichlet process [see, e.g., Green and Richardson (2001), or expand (7)]:

$$\frac{\Gamma(\alpha)}{\Gamma(n+\alpha)}\alpha^d \prod_{j=1}^{d}(n_j-1)!$$

is the probability of any particular partition of the $n$ items into $d$ clusters of sizes $n_1, n_2, \ldots, n_d$. (These clusters are of items (genes) identical by descent, or through the structural model of Section 5.2, not by the chance event of the same allele being drawn twice from the gene pool.) In our present context the Dirichlet concentration parameter $\alpha$ is $M$ or $(1-\theta)/\theta$, and $n$ is the number of genes drawn from the gene pool. So to take the example of $n=3$ founding genes, there are three possibilities: (a) all three are different; (b) two are the same, and the other different; (c) all three are the same. The respective probabilities are proportional to $(\alpha^2, 3\alpha, 2)$. If two modules analogous to that in Figure 5, with concentration parameters $\alpha$, $\beta$ respectively, are cascaded together, it is straightforward but tedious to verify that the probabilities of these three possibilities are now proportional to

$$(\alpha^2\beta^2, 3\alpha\beta(\alpha+\beta+2), (\alpha+\beta)(\alpha+\beta+3)-\alpha\beta+4),$$

which are not in the correct ratio. Thus, if, for example, we use (6) to model both uncertainty in allele frequencies and identity by descent, the pattern of dependence induced by the two factors combined is not properly described by (6) in any problem involving three or more founding genes.

However, it is true that to first order the Dirichlet modules do cascade together without changing their functional form. For large $\alpha$ and $\beta$, the probability that any particular pair of the $n$ genes are identical is $\alpha^{-1}+\beta^{-1}+O(\alpha^{-2}, \alpha^{-1}\beta^{-1}, \beta^{-2})$, the probability that all are distinct is $1-\binom{n}{2}(\alpha^{-1}+\beta^{-1})+O(\alpha^{-2}, \alpha^{-1}\beta^{-1}, \beta^{-2})$, and all other possibilities have negligible probability. Thus, for large $M$/small $\theta$, the effects of combining uncertainty in allele frequencies, identity by descent or population heterogeneity are additive on the scale of $M^{-1}=\theta/(1-\theta)$.

For exact results for specific models for the phenomena, the Bayes net modules for UAF, HET and IBD can be combined, properly reflecting the genetics of the situation. The only two realistic scenarios are UAF followed by HET and UAF followed by IBD. (Combinations of IBD and HET are unrealistic since two genes that might be IBD cannot belong to different subpopulations.) Thus output genes from the Pólya urn scheme representing uncertainty in allele frequencies can be fed into the model for heterogeneous populations; or are subject to patterned selection according to the IBD model.

There have been previous algebraic approaches for different scenarios combining multiple sources of dependence [see Ayres and Overall (1999), Weir (2007a), and Fung and Hu (2004), among others].

## 6. Results for the different scenarios on various forensic examples.

Throughout this section, when not stated otherwise, the following settings are used for illustrative purposes. For UAF, the aggregated prior and sample size $M$ is set to 100. For the IBD scenario we assume that two actors are either unrelated (with probability 0.90) or that the possible relation between them is equally likely to be that of parent–child or of half sibs (see the first and third block of rows of Table 12) with $\alpha = \gamma = 0.05$.

We need to clarify that the choice of the parameters and the possible relationships among actors yielding the numerical results given here are merely illustrative. The potential user of the methods we describe would need to insert his/her beliefs about the likely relationships. Here we have chosen the possibility of particular close relationships among actors.

The baseline model corresponds to $M \to \infty$ for UAF and $\alpha = \gamma = 0$ for IBD. The analyses for the baseline, the IBD and UAF scenarios are based on random draws from the Caucasian gene pool in Butler et al. (2003). For the HET scenario the $k$ actors are drawn from possibly different components $S_i$, $i = 1, \ldots, k$, of a mixed population which is an equal mixture of the Afro-Caribbean, Hispanic and Caucasian populations. Equal prior weights are used here purely for illustrative purposes, weights based on real population sizes or weights based on other non-DNA evidence should be used in real forensic casework.

### 6.1. *Criminal identification.*

Based on the evidence for the criminal identification case in Table 1, likelihood ratios were computed for the baseline assumptions and for UAF, IBD and HET scenarios; these are presented in columns 2 to 5 of Table 5. Both marker by marker and overall results are given, along with across-marker results using the product rule. The results for HET refer to the case when both the suspect `s`, and the unknown alternative suspect `as`, are from different components $S_1$ and $S_2$, of a heterogeneous mixed population.

Table 6 gives the allele frequencies for markers D3 and THO1 in the three subpopulations considered. Comparing UAF and IBD to the baseline scenario, one can see that both systematically down-weight the LR. The effect

TABLE 5

*Likelihood ratios for criminal identification for baseline, UAF, IBD and HET, as well as combinations of UAF plus IBD and UAF plus HET*

| Marker | Baseline | UAF | IBD | HET | UAF+IBD | UAF+HET |
|---|---|---|---|---|---|---|
| D13 | 138.9 | 106.6 | 88.7 | 126.7 | 71.7 | 113.9 |
| D3 | 1162.8 | 194.6 | 111.9 | 3488.4 | 74.3 | 583.7 |
| D5 | 27.7 | 23.6 | 20.5 | 35.6 | 18.2 | 33.4 |
| D7 | 16.9 | 14.6 | 13.7 | 11.8 | 12.1 | 11.2 |
| FGA | 12.3 | 11.8 | 11.1 | 17.0 | 10.6 | 16.7 |
| THO1 | 27.7 | 22.7 | 21.0 | 10.3 | 17.8 | 22.5 |
| TPOX | 36.7 | 31.5 | 27.5 | 35.8 | 24.3 | 34.2 |
| VWA | 25.0 | 20.8 | 19.2 | 32.2 | 16.5 | 29.0 |
| Overall $\log_{10}$LR | | | | | | |
| Exact | 13.38 | 12.10 | 7.71 | 13.85 | 7.49 | 12.57 |
| Product rule | 13.38 | 12.10 | 11.54 | 13.57 | 10.95 | 12.96 |

can be dramatic when the baseline frequency of an allele is very rare, as for marker D3. For $M = 10{,}000$ we attain near convergence to the baseline value (not shown here).

Furthermore, for UAF the product rule (2) applies and the overall LR is roughly 20 times smaller than the baseline. For IBD the evidence is roughly $460 \times 10^3$ times smaller than the baseline and the product rule overestimates the weight of evidence yielding a value about $7 \times 10^3$ times greater than the exact value.

It is important to observe that HET induces dependence among the markers; the product rule does not apply. Furthermore, as stated in Section 5.4, one would get a different result by using a gene pool of the weighted average subpopulation frequencies, instead of explicitly modeling the subpopulation mixture. For HET, however, there is no regular pattern in the LR values. For example, for marker D3 the LR for HET is more than 3 times that of the baseline using the Caucasian allele frequencies (allele 11 being absent in

TABLE 6

*Selected allele frequencies in the different subpopulations for markers D3 and THO1*

| Marker | Allele | Caucasian | Afro-Caribbean | Hispanic |
|---|---|---|---|---|
| D3 | 11 | 0.002 | 0 | 0 |
| | 17 | 0.215 | 0.205 | 0.204 |
| THO1 | 7 | 0.190 | 0.421 | 0.279 |
| Database size (individuals) | | 302 | 258 | 140 |

the other two subpopulations, see Table 6); whereas for marker THO1, the LR for HET is less than half the value for the baseline, since allele 7 has a much lower frequency in the Caucasian subpopulation than in the other two (again, see Table 6). In real forensic casework we would amend the values of the allele frequencies by updating the gene pool database with the newly observed individuals [Dawid and Mortera (1996)].

The last two columns of Table 5 show the LR for the combination of scenarios UAF plus IBD and UAF plus HET. The overall effect of combining UAF plus IBD is dramatic—it reduces the overall LR by a factor $775 \times 10^5$ compared to the baseline, and is roughly $3 \times 10^3$ times smaller than the product rule result. For the combination of UAF plus HET the overall LR is roughly 6 times smaller than the baseline and 2.5 times smaller than the product rule result.

An initial study of the usefulness of the analytic methods CSD and LFP of assessing sensitivity, discussed in Section 3.5 and Section 3.6, was conducted for each of the UAF and IBD scenarios. The linear constraints imposed for illustration restrict attention to joint distributions for the two genes of each founding actor that (a) are exchangeable with respect to actor, (b) are exchangeable with respect to gene within each actor, and (c) give the same marginal distribution for each gene as the baseline. Bounds were computed for departures from baseline that are both absolute and relative (that is, weighted by the baseline probabilities), in all cases separately for each marker.

For comparison with the exact results computed as discussed using BNs, relevant values for $\varepsilon$ were calculated from these exact results, thus, $\varepsilon_{\mathrm{A}}^{\mathrm{CSD}} = \|f - f_0\|$, $\varepsilon_{\mathrm{R}}^{\mathrm{CSD}} = \|f/f_0 - 1\|$ (with division defined componentwise), $\varepsilon_{\mathrm{A}}^{\mathrm{LFP}} = \max_{\mathbf{i}} |f_{\mathbf{i}} - f_{0\mathbf{i}}|$ and $\varepsilon_{\mathrm{R}}^{\mathrm{LFP}} = \max_{\mathbf{i}} |f_{\mathbf{i}} - f_{0\mathbf{i}}|/f_{0\mathbf{i}}$. Here $f$ denotes the joint distribution of the founding genes under the scenario in question, and $f_0$ the baseline distribution.

The resulting bounds for the likelihood ratios are presented in Table 7. The absolute bounds obtained from the linear fractional method are suppressed, since all are $(0, \infty)$. Trivial bounds of 0 and $\infty$ arise when the ellipsoids or rectangles defined by the values of $\varepsilon$ above include values for which marginal likelihoods are 0 under one or other hypothesis.

Our preliminary conclusion from these results is that these analytic methods do not provide bounds that are likely to be useful in forensic casework, for departures from the baseline as large as those considered here.

6.2. *Mixed trace.* The analyses of the different scenarios for the mixed trace example of Section 4.2 based on the evidence in Table 2 are shown in Table 8. Only the LR for hypothesis $H_0$: $s\&v$ against $H_1$: $as\&v$ is given. UAF refers to the case where all 8 founding genes `s`, `as`, `v` and `av` are uncertain, and HET refers to the case where `as` and `av`'s genes are drawn from possibly differing components of the mixed populations.

TABLE 7
*Criminal identification: bounds on likelihood ratios from constrained steepest descent (CSD) and linear fractional programming (LFP) analyses of UAF and IBD scenarios: † denotes "bounds" that fail to bracket the exact value*

| | | | **lfp, relative** | | **csd, relative** | | **csd, absolute** | |
|---|---|---|---|---|---|---|---|---|
| | **Baseline** | **Exact** | **Lower** | **Upper** | **Lower** | **Upper** | **Lower** | **Upper** |
| **UAF** | | | | | | | | |
| D13 | 138.9 | 106.6 | 0 | ∞ | 20.6 | ∞ | 41.4 | ∞ |
| D3 | 1162.8 | 194.6 | 0 | ∞ | 123.6 | ∞ | 1162.2† | ∞ |
| D5 | 27.7 | 23.6 | 0 | ∞ | 11.1 | ∞ | 24.8† | ∞ |
| D7 | 16.9 | 14.6 | 11.8 | 24.6 | 13.6 | 22.1 | 6.2 | ∞ |
| FGA | 12.3 | 11.8 | 7.4 | 21.1 | 9.7 | 16.8 | 5.0 | ∞ |
| THO1 | 27.7 | 22.7 | 16.4 | 47.4 | 20.1 | 42.0 | 5.1 | ∞ |
| TPOX | 36.7 | 31.5 | 0 | ∞ | 14.8 | 513.7 | 27.9 | ∞ |
| VWA | 25 | 20.8 | 15.3 | 41.2 | 18.5 | 36.5 | 5.3 | ∞ |
| **IBD** | | | | | | | | |
| D13 | 138.9 | 88.7 | 0 | ∞ | 13.5 | ∞ | 41.4 | ∞ |
| D3 | 1162.8 | 111.9 | 0 | ∞ | 123.6† | ∞ | 1162.2† | ∞ |
| D5 | 27.7 | 20.5 | 0 | ∞ | 9.4 | ∞ | 24.8† | ∞ |
| D7 | 16.9 | 13.7 | 10.9 | 26.7 | 12.4 | 25.6 | 4.8 | ∞ |
| FGA | 12.3 | 11.1 | 6.7 | 23.5 | 8.5 | 21.1 | 4.0 | ∞ |
| THO1 | 27.7 | 21 | 14.8 | 52.8 | 18.0 | 51.4 | 3.8 | ∞ |
| TPOX | 36.7 | 27.5 | 0 | ∞ | 11.9 | ∞ | 27.9† | ∞ |
| VWA | 25 | 19.2 | 13.9 | 45.6 | 16.7 | 44.1 | 4.0 | ∞ |

The baseline overall result is roughly 1.8, 55 and 1.2 times bigger than those for UAF, IBD and HET, respectively. The product rule yields an answer about 23 times bigger than the correct result for IBD; for HET it is about 1.2 times smaller.

Fung and Hu (2004) derive algebraic formulae for analyzing various mixed trace examples when considering specific relationships among actors as well as using the Balding and Nichols (1995) correction to allow for population structure. This example involves combining different scenarios, as in Section 5.5. We can derive all their results by considering the combination of UAF and a *fixed* relationship $R$ in the IBD model (not shown here). However, with our methodology we can extend their analysis to model uncertainty over $R$. Recall, that in this case one cannot simply obtain the overall LR by applying the product rule, so their algebraic method might become too difficult to implement.

6.3. *Simple paternity testing.* The overall LR for paternity testing based on the evidence in Table 3 shows less dramatic departures from baseline, as can be seen in Table 9. Column UAF1 refers to the case where only `pf` and `af` have uncertain allele frequencies and UAF2 refers to the case where all founders, `pf`, `af` and `m`, have uncertain allele frequencies. Recall, the results for HET correspond to the case when `pf` and `af` are drawn from different components, `S1` and `S2`, of the subpopulation. Again, the biggest difference in the results presented in Table 9 occurs between the baseline and the IBD scenario where the LR is roughly 6.5 times less. Furthermore, for IBD the LR for the product rule result is roughly 4 times bigger than the exact LR, whereas, for HET the product rule underestimates the exact LR by about 10%.

TABLE 8
*Comparison of likelihood ratios for a DNA mixture for baseline, UAF, IBD and HET scenarios*

| **Marker** | **Baseline** | **UAF** | **IBD** | **HET** |
|---|---|---|---|---|
| D13 | 5.22 | 4.85 | 4.83 | 7.17 |
| D3 | 7.10 | 6.38 | 6.22 | 6.72 |
| D5 | 3.63 | 3.36 | 3.40 | 3.53 |
| D7 | 4.86 | 4.68 | 4.53 | 3.97 |
| FGA | 51.78 | 46.17 | 39.02 | 34.94 |
| THO1 | 5.62 | 5.01 | 5.09 | 4.18 |
| TPOX | 3.13 | 3.10 | 3.00 | 3.47 |
| VWA | 6.56 | 6.18 | 6.01 | 8.44 |
| Overall $\log_{10}$LR | | | | |
| Exact | 6.59 | 6.33 | 4.85 | 6.52 |
| Product rule | 6.59 | 6.33 | 6.22 | 6.46 |

Table 10 gives further details on population heterogeneity when: (a) `pf` and `af` are drawn from different mixture components (the same as column HET in Table 9); (b)–(d) `pf` is drawn from the mixed population and `af` is drawn from a Caucasian, Afro-Caribbean and Hispanic gene pool, respectively. Note that the product rule applies for cases (b)–(d), where one of the untyped actors `af` is from a specified subpopulation. This will be true in general, whereas for (a), the product rule understates the overall weight of evidence.

TABLE 9
*Likelihood ratios for paternity testing for baseline UAF, IBD and HET*

| | **Baseline** | **UAF1** | **UAF2** | **IBD** | **HET** |
|---|---|---|---|---|---|
| Exact | 1317.56 | 1007.53 | 912.33 | 202.29 | 1313.32 |
| Product rule | 1317.56 | 1007.53 | 912.33 | 797.69 | 1209.60 |

TABLE 10
*Likelihood ratios for paternity testing for different subpopulation scenarios (`pf` is drawn from `S1`)*

| | (a) `af` from `S2` | (b) `af` Cauc. | (c) `af` Afro. | (d) `af` Hisp. |
|---|---|---|---|---|
| Exact | 1313.32 | 1317.56 | 1886.15 | 1004.90 |
| Product rule | 1209.60 | 1317.56 | 1886.15 | 1004.90 |

6.4. *Disputed sibship.* Table 11 shows results for the different scenarios for the disputed sibship case of Section 4.4, illustrated in Figure 4, based on the evidence in Table 4. For UAF1, only `af` and `tf2` are modeled as having uncertain allele frequencies, whereas in case UAF2 all 8 founders are modeled as having uncertain allele frequencies. Again, for HET `af` and `tf2` are drawn from possibly different subpopulation components.

Comparing UAF1 and UAF2, we note that the overall LR decreases slightly in the latter case.

In this example, based on indirect evidence—actor `tf2`'s DNA profile was not available—the LR is quite weak. For example, under a uniform prior probability on the target `tf1=tf2`, the LR yields a posterior probability of 0.747 for the baseline and 0.694 for UAF2.

**7. Conclusions and further work.** This paper illustrates some approaches for analyzing the sensitivity of forensic identification inference in Bayesian networks to assumptions about joint allele frequency distributions of founding genes. These are demonstrated on several different examples involving DNA evidence such as criminal identification, paternity testing, disputed sibship and mixed traces.

TABLE 11
*Likelihood ratios for disputed sibship*

| Marker | Baseline | UAF1 | UAF2 | IBD | HET |
|---|---|---|---|---|---|
| D13 | 4.032 | 3.806 | 3.681 | 3.621 | 3.876 |
| D3 | 0.354 | 0.353 | 0.352 | 0.362 | 0.356 |
| D5 | 2.120 | 2.083 | 2.024 | 2.034 | 2.369 |
| D7 | 0.402 | 0.401 | 0.395 | 0.411 | 0.387 |
| FGA | 0.444 | 0.441 | 0.443 | 0.453 | 0.480 |
| THO1 | 3.472 | 3.338 | 3.443 | 3.177 | 2.516 |
| TPOX | 0.473 | 0.470 | 0.467 | 0.483 | 0.450 |
| VWA | 3.333 | 3.212 | 3.084 | 3.065 | 3.711 |
| Overall LR | | | | | |
| Exact | 2.956 | 2.490 | 2.273 | 2.285 | 2.501 |
| Product rule | 2.956 | 2.490 | 2.273 | 2.341 | 2.552 |

A first approach assesses the sensitivity of the inference to founders by building up and running a Bayesian network for a variety of different scenarios of interest in forensic genetics, such as uncertainty in allele frequencies (UAF), identity by descent (IBD) and in the presence of population heterogeneity (HET), as well as plausible combinations of these scenarios. We show that IBD and HET scenarios induce dependence among markers which need to be handled simultaneously, since the simplifying product rule no longer applies. Here we have also attempted to clarify the relation between the standard approach using the Balding and Nichols correction formula and the model we use for uncertain allele frequency.

A second approach is based on analytic methods including constrained steepest descent (CSD) and linear fractional programming (LFP). Results using this approach can give bounds which are rather wide.

In casework analysis, including the possibility of UAF, IBD, HET and plausible combinations thereof could transform a LR that is incriminating in the baseline scenario to one that is below the threshold for incriminating a suspect or declaring paternity. It is important in cases brought before the court to present a result that errs on the side of caution, that is, which is less incriminating for the suspect. When incorporating UAF and IBD and in some HET scenarios, the correct LRs are less incriminating than those computed naively assuming standard assumptions. In all the examples analyzed here the *product rule* consistently overestimates the LR for the IBD scenario.

The numerical results presented are computed in both GRAPPA, and with the HUGIN software, available at `http://www.hugin.com`. Codes used in our examples, for both of these systems, can be found at http://www.stats.bris.ac.uk/~peter/Sensitivity.

The modularity and flexibility of the approach based on Bayesian networks allows ready application to numerous different examples and complicating features that have not been analyzed here. For example, the deconvolution of DNA mixed traces using quantitative peak area information has been solved using Bayesian networks [Cowell, Lauritzen and Mortera (2007a, 2007b)]. Further forensic genetics applications handled using Bayesian networks account for the possibility of mutation, as well as artifacts such as allelic drop-out and the presence of silent alleles [Dawid, Mortera and Vicard (2007)]. The effect on the LR of introducing mutation and silent alleles can also be substantial even when the underlying perturbations are small. It should be reasonably straightforward to incorporate the basic modular scenarios described here, in these examples as well. A further application of our methods for incorporating IBD, to assess sensitivity to quantities other than the target hypotheses like guilt/innocence, could be that of inferring the posterior probability of specific relationships among actors conditional on their DNA profiles [Egeland et al. (2000)]. This could be useful, for example, in immigration cases.

The methodology developed could have a much wider applicability than forensic genetics applications. For example, the UAF scenario could be used for modeling the uncertainty in name distributions used in the identification of archaeological finds [Mortera and Vicard (2008)].

## APPENDIX

**A.1. Uncertain allele frequencies.** Pseudo-code for Pólya urn representation. `GP()` denotes a draw from the gene pool, and `Bernoulli(`$p$`)` a draw from Bernoulli ($p$):

```
g[1]~GP()
for(i in 2:N) pool[i]~GP()
for(i in 2:N) c[i]~Bernoulli(M/(M+i-1))
g[2] = if c[2] then pool[2] else g[1]
for(i in 3:N) {
for(j in 2:(i-1))
     d[i,j]~Bernoulli(1/j)
     temp[i,2] = if d[i,2] then g[2] else g[1]
     if(i>3) for(j in 3:(i-1))
      temp[i,j] = if d[i,j] then g[j] else temp[i,j-1]
     g[i] = if c[i] then pool[i] else temp[i,i-1]
}
```

**A.2. Identity by descent.** The joint distribution of patterns of IBD between the 2 genes of 2 actors with 9 different degrees of relatedness $R$, each one, where applicable, treated symmetrically over the two actors and two sexes (e.g., parent-child has 4 arrangements that matter—`a` father to `b`, `b` mother to `a`, etc.) is given in Table 12 and in Table 13 (for incestuous relationships). It has a panel (separated by white space) for each relationship $R$; the 2nd column is the assumed probability of the relationship, the 3rd column the probability, given the relationship, of the pattern $\pi_m$ of IBD expressed by the indicators in the remaining columns. For each relationship, there should be a line for no IBD (all zero indicators)—it is omitted.

Summarizing the probabilities across all possible relationships given in Table 12 and Table 13, we have the following:

(a) `apg=bpg` others differ OR `amg=bmg` others differ

$$\alpha/4 + \beta/4 + \gamma/4 + \delta/8 + \varepsilon/16 + 3\phi/32 + \psi/64 + \lambda/8 + \mu/8,$$

(b) `apg=bmg` others differ OR `amg=bpg` others differ

$$\alpha/4 + \delta/8 + \varepsilon/16 + 3\phi/32 + \psi/64 + \lambda/8 + \mu/32,$$

Table 12
*Patterns of dependence for IBD*

| $R$ | $p(R)$ | $p(\pi_m\|R)$ | apg=amg | apg=bpg | apg=bmg | amg=bpg | amg=bmg | bpg=bmg | # common |
|---|---|---|---|---|---|---|---|---|---|
| a father of b | $\alpha/4$ | 0.5 | 0 | 1 | 0 | 0 | 0 | 0 | 1 |
| | | 0.5 | 0 | 0 | 0 | 1 | 0 | 0 | 1 |
| a mother of b | $\alpha/4$ | 0.5 | 0 | 0 | 1 | 0 | 0 | 0 | 1 |
| | | 0.5 | 0 | 0 | 0 | 0 | 1 | 0 | 1 |
| b father of a | $\alpha/4$ | 0.5 | 0 | 1 | 0 | 0 | 0 | 0 | 1 |
| | | 0.5 | 0 | 0 | 1 | 0 | 0 | 0 | 1 |
| b mother of a | $\alpha/4$ | 0.5 | 0 | 0 | 0 | 1 | 0 | 0 | 1 |
| | | 0.5 | 0 | 0 | 0 | 0 | 1 | 0 | 1 |
| Sibs | $\beta$ | 0.25 | 0 | 1 | 0 | 0 | 1 | 0 | 2 |
| | | 0.25 | 0 | 1 | 0 | 0 | 0 | 0 | 1 |
| | | 0.25 | 0 | 0 | 0 | 0 | 1 | 0 | 1 |
| 1/2 sibs, same mother | $\gamma/2$ | 0.5 | 0 | 0 | 0 | 0 | 1 | 0 | 1 |
| 1/2 sibs, same father | $\gamma/2$ | 0.5 | 0 | 1 | 0 | 0 | 0 | 0 | 1 |
| a sib of father of b | $\delta/4$ | 0.25 | 0 | 1 | 0 | 0 | 0 | 0 | 1 |
| | | 0.25 | 0 | 0 | 0 | 1 | 0 | 0 | 1 |
| a sib of mother of b | $\delta/4$ | 0.25 | 0 | 0 | 1 | 0 | 0 | 0 | 1 |
| | | 0.25 | 0 | 0 | 0 | 0 | 1 | 0 | 1 |
| b sib of father of a | $\delta/4$ | 0.25 | 0 | 1 | 0 | 0 | 0 | 0 | 1 |
| | | 0.25 | 0 | 0 | 1 | 0 | 0 | 0 | 1 |
| b sib of mother of a | $\delta/4$ | 0.25 | 0 | 0 | 0 | 1 | 0 | 0 | 1 |
| | | 0.25 | 0 | 0 | 0 | 0 | 1 | 0 | 1 |
| Cousins, mothers are sibs | $\varepsilon/4$ | 0.25 | 0 | 0 | 0 | 0 | 1 | 0 | 1 |
| Cousins, mother a sib of father b | $\varepsilon/4$ | 0.25 | 0 | 0 | 0 | 1 | 0 | 0 | 1 |
| Cousins, father a sib of mother b | $\varepsilon/4$ | 0.25 | 0 | 0 | 1 | 0 | 0 | 0 | 1 |
| Cousins, fathers are sibs | $\varepsilon/4$ | 0.25 | 0 | 1 | 0 | 0 | 0 | 0 | 1 |
| Double cousins, same sex parents sibs | $\phi/2$ | 0.0625 | 0 | 1 | 0 | 0 | 1 | 0 | 2 |
| | | 0.1875 | 0 | 0 | 0 | 0 | 1 | 0 | 1 |
| Double cousins, opposite sex parents sibs | $\phi/2$ | 0.0625 | 0 | 0 | 1 | 1 | 0 | 0 | 2 |
| | | 0.1875 | 0 | 0 | 0 | 1 | 0 | 0 | 1 |
| 2nd cousins, mothers are cousins | $\psi/4$ | 0.0625 | 0 | 0 | 0 | 0 | 1 | 0 | 1 |
| 2nd cousins, mother a cousin of father b | $\psi/4$ | 0.0625 | 0 | 0 | 0 | 1 | 0 | 0 | 1 |
| 2nd cousins, father a cousin of mother b | $\psi/4$ | 0.0625 | 0 | 0 | 1 | 0 | 0 | 0 | 1 |
| 2nd cousins, fathers are cousins | $\psi/4$ | 0.0625 | 0 | 1 | 0 | 0 | 0 | 0 | 1 |

Table 13
*Patterns of dependence for IBD for incestuous scenarios*

| $R$ | $p(R)$ | $p(\pi_m\|R)$ | apg=amg | apg=bpg | apg=bmg | amg=bpg | amg=bmg | bpg=bmg | # common |
|---|---|---|---|---|---|---|---|---|---|
| b mother and sister to a | $\lambda/4$ | 0.25 | 1 | 1 | 0 | 1 | 0 | 0 | 3 |
| | | 0.25 | 0 | 0 | 0 | 1 | 0 | 0 | 1 |
| | | 0.25 | 0 | 1 | 0 | 0 | 1 | 0 | 2 |
| | | 0.25 | 0 | 0 | 0 | 0 | 1 | 0 | 1 |
| b father and brother to a | $\lambda/4$ | 0.25 | 1 | 0 | 1 | 0 | 1 | 0 | 3 |
| | | 0.25 | 0 | 0 | 1 | 0 | 0 | 0 | 1 |
| | | 0.25 | 0 | 1 | 0 | 0 | 1 | 0 | 2 |
| | | 0.25 | 0 | 1 | 0 | 0 | 0 | 0 | 1 |
| a mother and sister to b | $\lambda/4$ | 0.25 | 0 | 1 | 1 | 0 | 0 | 1 | 3 |
| | | 0.25 | 0 | 0 | 1 | 0 | 0 | 0 | 1 |
| | | 0.25 | 0 | 1 | 0 | 0 | 1 | 0 | 2 |
| | | 0.25 | 0 | 0 | 0 | 0 | 1 | 0 | 1 |
| a father and brother to b | $\lambda/4$ | 0.25 | 0 | 0 | 0 | 1 | 1 | 1 | 3 |
| | | 0.25 | 0 | 0 | 0 | 1 | 0 | 0 | 1 |
| | | 0.25 | 0 | 1 | 0 | 0 | 1 | 0 | 2 |
| | | 0.25 | 0 | 1 | 0 | 0 | 0 | 0 | 1 |
| Parents are sibs | $\mu$ | 0.0625 | 1 | 1 | 1 | 1 | 1 | 1 | 4 |
| | | 0.0625 | 0 | 0 | 0 | 1 | 1 | 1 | 3 |
| | | 0.0625 | 0 | 1 | 1 | 0 | 0 | 1 | 3 |
| | | 0.0625 | 1 | 0 | 1 | 0 | 1 | 0 | 3 |
| | | 0.0625 | 1 | 1 | 0 | 1 | 0 | 0 | 3 |
| | | 0.03125 | 0 | 0 | 1 | 1 | 0 | 0 | 2 |
| | | 0.03125 | 1 | 0 | 0 | 0 | 0 | 1 | 2 |
| | | 0.1875 | 0 | 1 | 0 | 0 | 1 | 0 | 2 |
| | | 0.125 | 0 | 0 | 0 | 0 | 1 | 0 | 1 |
| | | 0.125 | 0 | 1 | 0 | 0 | 0 | 0 | 1 |
| | | 0.03125 | 0 | 0 | 0 | 1 | 0 | 0 | 1 |
| | | 0.03125 | 0 | 0 | 1 | 0 | 0 | 0 | 1 |
| | | 0.03125 | 1 | 0 | 0 | 0 | 0 | 0 | 1 |
| | | 0.03125 | 0 | 0 | 0 | 0 | 0 | 1 | 1 |

(c) `apg=amg` others differ OR `bpg=bmg` others differ

$$\mu/32,$$

(d) `apg=bpg` AND `amg=bmg`, but these differ

$$\beta/4+\phi/32+\lambda/4+3\mu/16,$$

(e) `apg=bmg` AND `amg=bpg`, but these differ

$$\phi/32+\mu/32,$$

(f) apg=bpg AND amg=bpg, but these differ

$$\mu/32,$$

(g) all same except apg OR all same except amg OR all same except bpg OR all same except bmg

$$\lambda/16 + \mu/16,$$

(h) all same

$$\mu/16.$$

This yields a total of

$$\alpha + 3\beta/4 + \gamma/2 + \delta/2 + \varepsilon/4 + 7\phi/16 + \psi/16 + \lambda + 15\mu/16.$$

The text above summarizes the total probability across the scenarios. All possible patterns appear. The only symmetries arise from switching the actors, or switching the sexes. For the range of relationships considered here one can derive inequality constraints such as

$$P(\texttt{apg=amg}) < P(\texttt{amg=bpg}) < P(\texttt{apg=bpg})$$

and symmetrically for

$$P(\texttt{bpg=bmg}) < P(\texttt{apg=bmg}) < P(\texttt{amg=bmg}).$$

Table 1 in Balding and Nichols (1995) refers to the numbers of genes that are IBD, not which genes they are, for 4 of the 9 degrees of relatedness given in Table 12 and Table 13.

**A.3. Software for Bayesian networks.** Many software systems, commercial or public-domain, are available for computations in Bayesian networks/probabilistic expert systems, and some of these will be suitable for the network calculations needed in the methods we discuss. These calculations do demand a degree of "programmability," so that looping over markers, etc., is straightforward, and this is not available in some systems.

Our numerical results have been obtained using two systems that do offer the necessary flexibility—HUGIN and GRAPPA.

HUGIN (http://www.hugin.com/) is a sophisticated commercial system for probabilistic networks. In recent editions, it implements the idea of Object Orientated Bayesian networks, which allows building networks from modules, which can be conveniently replicated and combined using a graphical interface. The networks illustrated in all of our figures are in fact screenshots of HUGIN network modules.

GRAPPA is a suite of functions in the statistical language R [R Development Core Team (2005)] that allows the construction of discrete

Bayesian networks on a modest scale, and inference in such models. It is freely available from http://www.stats.bris.ac.uk/~peter/Grappa. One of the advantages of its implementation in R is that all of the programming features of that language can be brought to bear to support the customization of Grappa to the problem at hand, and flexibility in experimentation. In particular, this extends to combining network computations with other kinds of programming, including the computations for the constrained steepest descent and linear fractional programming methods seen in Section 3.5 and Section 3.6.

Selected Grappa codes and Hugin networks used for our computations are available at http://www.stats.bris.ac.uk/~peter/Sensitivity.

School of Mathematics
University of Bristol
Bristol BS8 1TW
UK
E-mail: P.J.Green@bristol.ac.uk

Dipartimento di Economia
Università Roma Tre
Via S. D'Amico 77
00145 Roma
E-mail: mortera@uniroma3.it